\documentclass[amsmath,amssymb,nofootinbib]{revtex4}

\usepackage{xcolor}
\usepackage{latexsym,comment,verbatim}
\usepackage{amssymb}
\usepackage{amsfonts}
\usepackage{amsmath,color}
\usepackage{graphicx}
\usepackage{bm}
\usepackage{appendix}
\usepackage{pdfpages}
\usepackage{placeins}
\usepackage{caption}
\usepackage{subcaption}

\date{\today}

\begin{document}

\title{Fringe Contrast characterization and optimization for Ring Laser Gyroscopes}

\author{Nicolò Beverini$^1$, Giorgio Carelli$^{1,2}$, Simone Castellano$^{*,2,3}$, Giuseppe Di Somma$^{1,2}$, Angela D.V. Di Virgilio$^2$, Enrico Maccioni$^{1,2}$, Paolo Marsili$^{1,2}$}

\address{ 
$^1$ Dipartimento di Fisica, Universit\`a di Pisa, largo B. Pontecorvo 3, I-56127 Pisa, Italy \\
$^2$ Istituto Nazionale di Fisica Nucleare (INFN), sez. di Pisa, largo B. Pontecorvo 3, I-56127 Pisa, Italy \\
$^3$ Gran Sasso Science Institute, Viale Francesco Crispi 7, 67100 L'Aquila AQ, Italy
}

\email{simone.castellano@gssi.it, simone.castellano@pi.infn.it}

\begin{abstract}
Top  sensitivity Ring Laser Gyroscopes based on a simple mechanical scheme, so called heterolithic, have shown to operate on a  continuous basis; since the mirror positions are not fixed, unwanted signals can occur, and data selection is necessary. For this purpose, Fringe Contrast is very useful. It depends on output beams intensity, alignment, and polarization; the model of the alignment and polarization contributions to Contrast has been implemented, and tested on the GP2 prototype. Polarization changes are corrected by acting on the output beams with plates, in order to recombine linearly polarized light. Using such model, the GP2 prototype was characterized as for polarization of the beams,
inside the cavity and at their interference, and for non-planarity;
as a result, the interfering beams off-axis has also been measured.
\end{abstract}
\maketitle
\section{Introduction: the Ring Laser Gyroscope and the "Sagnac" signal}

Large area Ring Laser Gyroscopes (RLGs) have the capability of measuring the frame angular velocity in a large bandwidth, with unprecedented sensitivity. In general, they are based on a square cavity with side longer than 1m, currently the top sensitivity ones have a side around 4m long. 
The kind of measurement that can be performed with such devices is relevant in many scientific fields, from seismology \cite{lavoro_sismologia, belfi2017, igel2021, simonelli2016, simonelli2021} to fundamental physics, such as testing gravitomagnetic effects foreseen by general relativity, (i.e. the Lense Thirring effect), and beyond \cite{fundamental, altucci2023status, altucci2023memocs, capozziello2021, giovinetti2024}.

A RLG schematic view is shown in Fig.\ref{fig:rlg}.
Its working principle is based on the Sagnac effect. When the optical cavity of a ring laser is rotating around its axis, the time of round trip of the light beam co-rotating  with the cavity is slightly increased, while that of the counter-rotating one is slightly decreased. Then, a small frequency difference is produced between the  two  beams traveling in the opposite directions. 
This frequency difference, named  Sagnac frequency ($f_s$), is proportional to the frame rotational velocity $\Omega$ and can be observed  as the beat note between the two beams on a photo-detector, its derivation can be found in \cite{giovinetti2024}:

\begin{figure}[htbp]
    \centering
    \includegraphics[scale=0.5]{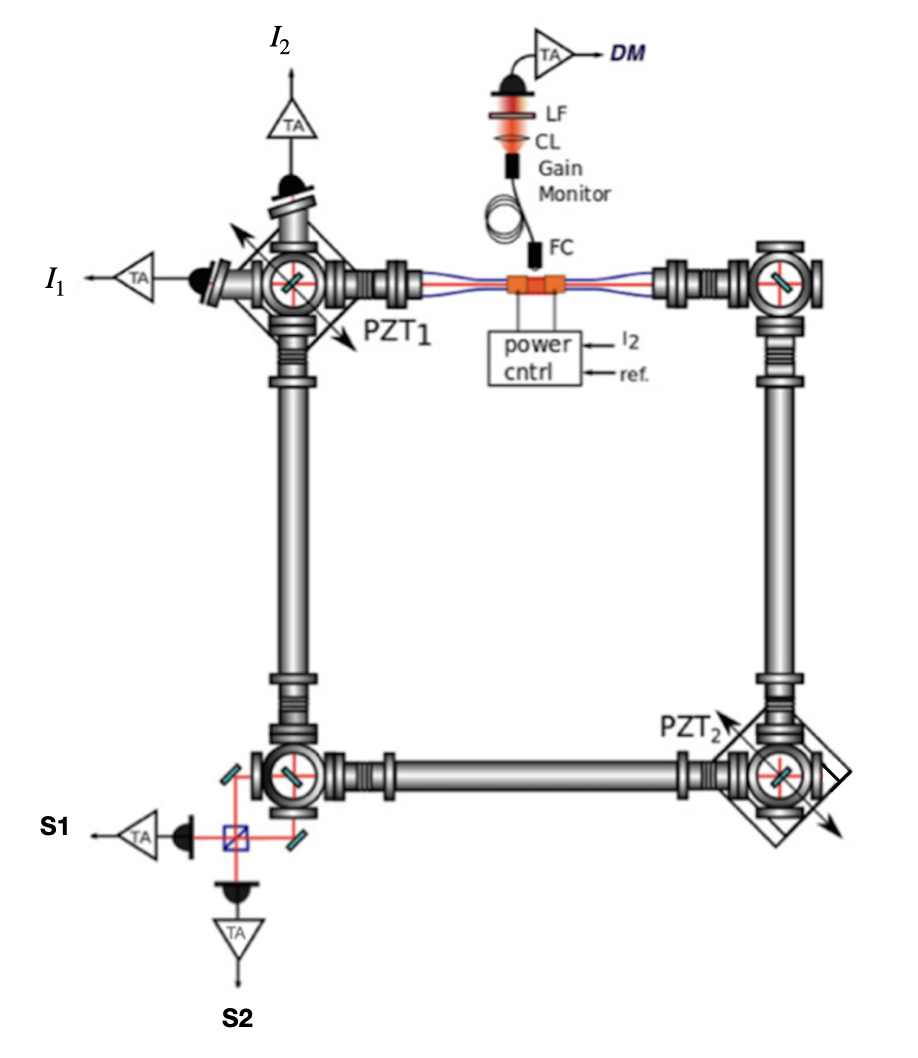}
    \caption{Schematic view of a typical Ring Laser Gyroscope (RLG). In particular, data acquired are the two laser beam intensities (I$_{1,2}$), and two beat note signals (S1,2). Two piezoelectric actuators (PZT$_{1,2}$) are placed for nano-metric tilt movements of the mirrors. The power control and Gain monitor provide feedback and stabilization of I$_1$, with respect to a reference level.}
\label{fig:rlg}
\end{figure}

\begin{equation}
   f_s = \frac{4A}{\lambda L} \hspace{1 mm} \Omega \hspace{1 mm} cos(\theta) = S \hspace{1 mm} \Omega \hspace{1 mm} cos(\theta),
   \label{scale_factor}
\end{equation}

where A is the area defined by the ring cavity, L is its perimeter, $\lambda$ the wavelength of the light, and $\theta$ is the angle between the area vector of the ring and the rotational axis. For a RLG optical resonator lying horizontally (area vector vertical) $\theta$ is the colatitude angle, while for RLGs aligned at the maximum Sagnac frequency $\theta$ = 0.
The frequency-lock of the two counter-propagating beams at low rotation rate of the optical cavity is a well known problem of RLGs; an improvement on the mirrors quality has allowed the use of the Earth rotation rate as a bias avoiding the frequency locking, ensuring continuous operation with the beams in slightly separated frequency.
The most performant RLG, operating in the geodetic observatory of Wetzell in Bavaria \cite{wetzell}, is monolithic, i.e. the mirrors are rigidly attached to a single piece of very low expansion material (ZERODUR); unfortunately, this is a costly design scheme, which poses several problems in the cavity orientation, optical alignment, and mirror changes.
RLG prototypes of the GINGER collaboration are based on  square optical cavities realised by enclosing the mirrors under vacuum inside metallic boxes,  guaranteeing the rigidity of the cavity frame with granite support and eventually actively controlling the geometry \cite{articolo_recente_enrico}. Since they are realised using pieces of different materials, they are called heterolithic (HL), a solution certainly advantageous for the cavity alignment.

The drawback of HL devices is that the mirrors are not perfectly fixed, and constant monitoring of the beams alignment and of the data quality is required; for this purpose, the Fringe Contrast model has been developed. 
Data are acquired from transmitted radiation, collecting the intensities of the two counter-propagating beams ($I_1$ and $I_2$ from here on), and of the two beat note signals (S1 and S2 from here on) obtained by the interference of the two beams exiting one of the corners and recombined at a beam-splitter.
The beat note signal is of the form $S=A\cdot cos(\omega t) + \zeta$, where $\omega$ is the pulsation and $\zeta$ is locally the mean value of S; S is a positive quantity, being physically a light intensity.
The reconstruction is performed by means of the Hilbert transform, which identifies amplitude and phase of the signal. By time deriving the phase term, and multiplying it for the acquisition rate, the measured beat note pulsation ($\omega_m$ from here on) is obtained. $\omega_m$ is however affected by laser non linear dynamics and some corrections must be done, in order to obtain the true value of $\omega_s = 2 \pi f_s$, as reported in \cite{virgilio2019,virgilio2020}.

A typical perturbation of HL operation is due to ’mode jumps’. The free spectral range of the optical cavity of a large frame RLG is quite small, competitive with the homogeneous width of the laser transition. Then, small perturbations (i.e. a change of the local temperature) can produce a sudden jump of the longitudinal operating order of the laser. In the data records, very fast spikes that affect data for a few seconds appear, and perturb the measurement of $\Omega$ at the level of a few nrad/s. Occasionally, the laser can enter in the regime of ’split mode’, where the two counter-propagating laser beams oscillate on two different longitudinal orders. These disturbances may last for hours and they impair the measurement of $\Omega$ for their whole duration. Typically, about 10$\%$ of the data must be removed in a free running device. However, when the geometry is controlled, the duty cycle is $ \sim  100\%$\cite{articolo_recente_enrico}.

In general, active control requires periodic checks, and it is convenient to leave the RLG free running, when continuous and long data taking is required, as for instance for geophysical applications, and when the device is operating in a distant experimental site.
A good indicator of data quality is Fringe Contrast (we will simply refer to it as C),
defined as:

\begin{equation}
    C = \frac{I_{max}-I_{min}}{I_{max}+I_{min}},
    \label{contrast}
\end{equation}
where "$I_{max}$" and "$I_{min}$" are the local intensity maxima and minima of the beat note signal.

\begin{figure}[!ht]
    \centering
    \includegraphics[width=6.0in]{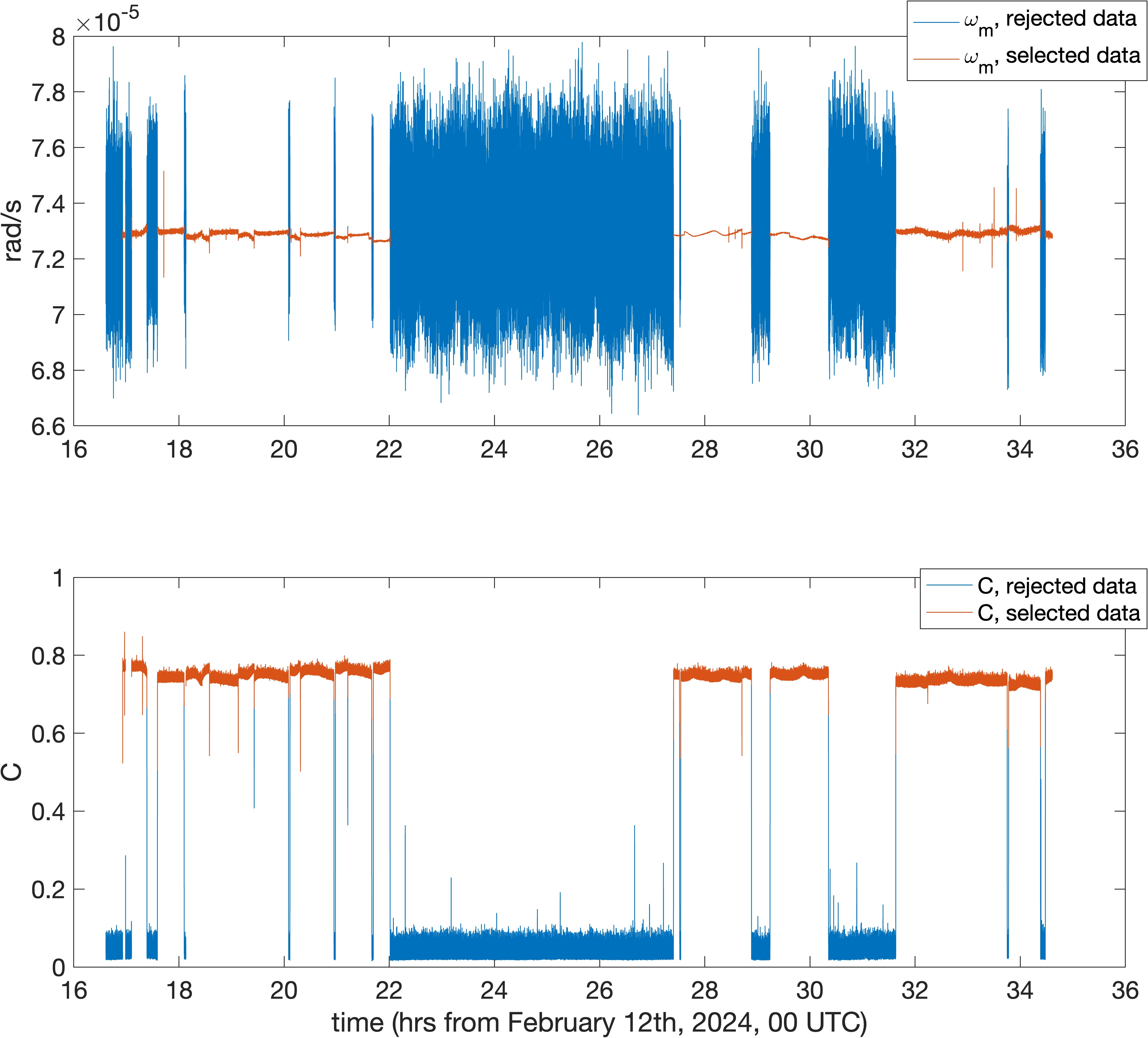}
    \caption{$\Omega$ obtained from $\omega_m$ (above), and C (below); how the selection is performed on the basis of C is shown.}
    \label{fig:raw_data}
\end{figure}

In Fig.\ref{fig:raw_data}, we show $\Omega$ obtained from the "raw" frequency $\omega_m$ (above), and C (below). Data shown and the tests described in the present paper are done using GP2, a HL RLG, 1.6m side, operating at the maximum Sagnac frequency, i.e. aligned with the Earth rotation axis.

Since $\zeta$ changes with time, it is a source of noise in the frequency reconstruction, and we have:

\begin{equation}
    \zeta = \frac{I_{max}-I_{min}}{2} = \frac{A}{2C},
    \label{zeta}
\end{equation}

from the definition of C. A higher C implies a better signal-to-noise ratio, therefore the scope of the present work, in optimizing C in a HL RLG.

The GP2 and GINGERino RLG prototypes have been recently both equipped with a double beat-note signal detection, and a $\sqrt{2}$ improvement of the  signal-to-noise ratio \cite{virgilio2019,virgilio2020,divirgilio2024} has been demonstrated. Within such efforts, a detailed study of the beams alignment and polarization was conducted, and the full characterization of a RLG prototype.

In the present article, we make a characterization of GP2 at INFN-Pisa, and show the improvement obtained by optimizing the beams output.

The work is structured as follows: in section 2, the implemented model of fringe Contrast, as function of the beams geometry and polarization, is described in detail. In section 3, we describe the beams polarization measurement for GP2. In section 4, the improvement that is obtained by making the beams linearly polarized at the cavity output is outlined, and a measurement of the GP2 beams off-axis is described.

\section{A model of the fringe Contrast}
From light interference, we know that local maxima and minima can be defined as:

\begin{equation}
    I_{max} = I_1 + I_2 + 2 \cdot \sqrt{I_1 \cdot I_2} \cdot \eta_{vis}
    \label{I_max}
\end{equation}

\begin{equation}
    I_{min} = I_1 + I_2 - 2 \cdot \sqrt{I_1 \cdot I_2} \cdot \eta_{vis},
    \label{I_min}
\end{equation}

where $\eta_{vis}$ is a coefficient called "visibility";
therefore:

\begin{equation}
    C = \frac{I_{max} - I_{min}}{I_{max} + I_{min}} =
    \frac{2 \cdot \sqrt{I_1 \cdot I_2}}{I_1 + I_2} \cdot \eta_{vis}
    \label{contrast_factors}
\end{equation}

We can factorize $\eta_{vis}$ as:

\begin{equation}
    \eta_{vis} = \eta_{geom} \cdot \eta_{pol},
    \label{eta_vis}
\end{equation}

where $\eta_{geom}$ is a contribution arising from the overlapping of the two beams at the point of interference, and $\eta_{pol}$ is the contribution arising from the polarizations of the two interfering beams.

\subsection{The beams alignment}
$\eta_{geom}$ was derived by considering the interference of two gaussian beams, of identical intensities ($I_1 = I_2 =1$) and geometry, propagating in the same z direction, with their z axes positioned at $y=0$, $x= \pm k$, being $x$ and $y$ the other two cartesian coordinates, perpendicular to the direction of the beams propagation.
The parameter $k$ is the distance between the two beams axes divided by the width of the gaussian beam profiles at $1/e^2$ in intensity.
$\eta_{pol}$ is set to 1, to consider only the geometrical contribution.
In such conditions, we have:

\begin{equation}
    \eta_{geom} = C = \frac{I_{max}-I_{min}}{I_{max}+I_{min}}
    \label{eta_geome}
\end{equation}

\begin{figure}[htbp]
    \centering
    \includegraphics[width=4.0in]{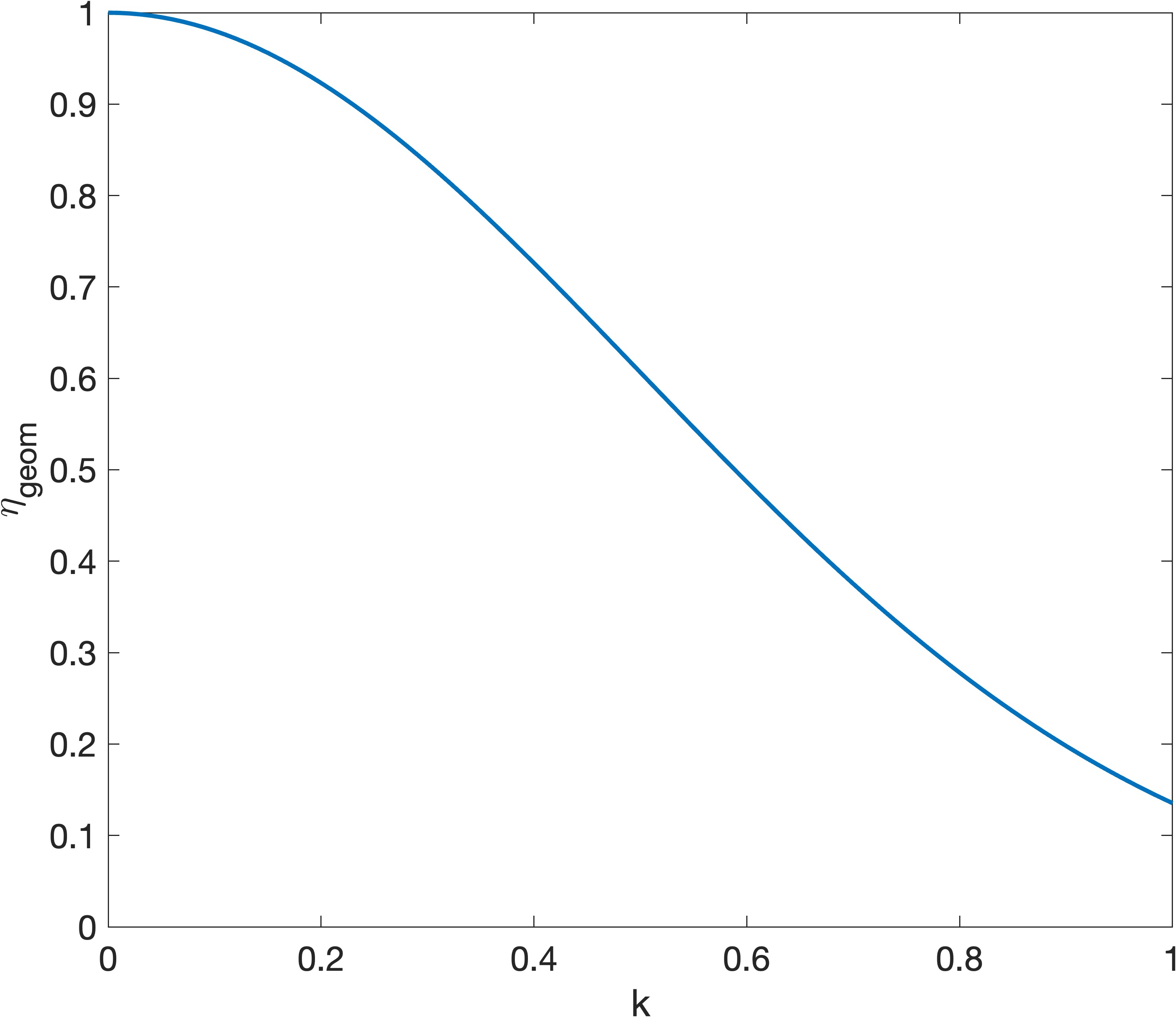}
    \caption{$\eta_{geom}$ expressed as a function of $k$.}
    \label{fig:eta_geome}
\end{figure}

Intensity maxima and minima ($I_{max}$ and $I_{min}$) of the beams interference were calculated as functions of $k$, therefore $\eta_{geom}$ was derived.
$\eta_{geom}$ as function of $k$ is shown in Fig.\ref{fig:eta_geome}.

\begin{equation}
    I_{max}= \frac{1}{2 \pi}\int_{-\infty}^{+\infty} \int_{-\infty}^{+\infty} (e^{-[(x+k)^2 + y^2]} + e^{-[(x-k)^2 + y^2]})^2 \,dx dy
    \label{i_min}
\end{equation}

\begin{equation}
    I_{min}= \frac{1}{2 \pi}\int_{-\infty}^{+\infty} \int_{-\infty}^{+\infty} (e^{-[(x+k)^2 + y^2]} - e^{-[(x-k)^2 + y^2]})^2 \,dx dy
    \label{i_max}
\end{equation}

From equations \ref{eta_geome}, \ref{i_min}, and \ref{i_max}, we derive:

\begin{equation}
    \eta_{geom} = e^{-2k^2}
    \label{eta_geome_a}
\end{equation}

\subsection{The beams polarization}
The most general situation that can we can modelize in terms of polarization consists in two beams exiting the cavity, with elliptical polarizations, and a certain angle between the ellipses axes.
The polarizations of the beams are elliptical if the cavity is not perfectly planar, and their ellipticity is connected to such non-planarity, as described in \cite{stedman90}. 
The angle (let's call it $\beta$) between the two polarization ellipses minor axes' (see Fig.\ref{fig:pol_rotation}) could instead be due to a possible birefringence of the output mirror. 
Hence, we calculate the contribution to contrast caused by the scenario of the two polarizations, $\eta_{pol}$, starting from the calculations in \cite{articolo_polarizzazioni}, and considering the same direction of propagation for the two interfering beams.
We remind that, for circular polarizations, it is, considering left-handed photons, without loss of generality:

\begin{figure}[htbp]
    \centering
    \includegraphics[scale=0.5]{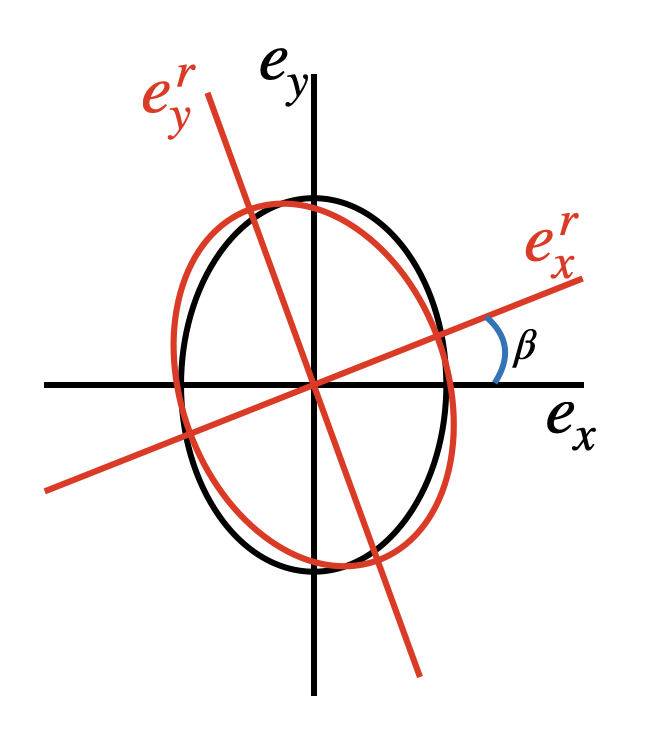}
    \caption{Schematic representation of a possible rotation of the elliptical polarization, between the two interfering beams.}
    \label{fig:pol_rotation}
\end{figure}

\begin{equation}
    \eta_{pol} = |e_L \cdot e_L^*| = 
    \frac{1}{2}|(e_x +ie_y)\cdot(e_x-ie_y)|=
    \frac{1}{2}(e_x^2+e_y^2)=1
    \label{eta_pol_circ}
\end{equation}

From the description of circular polarizations we can derive the description of elliptical polarizations, by introducing two different coefficients "a" and "b" that respectively multiply $e_x$ and $e_y$, they represent the polarization ellipses' semi-axes; if the angle is null between the two ellipses minor axes, we have:

\begin{equation}
    \eta_{pol} = 
    \frac{1}{2}|(ae_x +ibe_y)\cdot(ae_x-ibe_y)|=
    \frac{1}{2}(a^2e_x^2+b^2e_y^2)= \frac{1}{2}(a^2+b^2)=1,
    \label{eta_pol_ellipse}
\end{equation}

which gives us a normalization condition for the semi-axes of a beam polarization ellipse.
Now, if we introduce a non-zero angle between the axes of the ellipses that represent the polarizations of the counter-propagating beams, we can calculate again the  contribution of polarizations to contrast; if we rotate the polarization ellipse of an angle $\beta$, in one of the two interfering beams, we can refer to the orthogonal versors $e_x^r$ and $e_y^r$ (Fig.\ref{fig:pol_rotation}), rotated of an angle $\beta$ with respect to $e_x$ and $e_y$, respectively; we have:

\begin{equation}
\begin{split}
\eta_{pol} =
\frac{1}{2}|(ae_x +ibe_y) \cdot (ae_x^r-ibe_y^r)|=  \\
\frac{1}{2}|a^2(e_x \cdot e_x^r) + b^2 ( e_y \cdot e_y^r) + iab ( e_y \cdot e_x^r) - iab ( e_x \cdot e_y^r)|=  \\
\frac{1}{2}|(a^2cos(\beta) + b^2cos(\beta) + iab \cdot cos(\frac{\pi}{2}-\beta) - iab \cdot cos(\frac{\pi}{2}+\beta))|=  \\
|\frac{1}{2}(a^2+b^2)cos(\beta) + iab \cdot sin(\beta)|= \\
\sqrt{\frac{1}{4}(a^2+b^2)^2\cdot cos^2(\beta)+a^2b^2\cdot sin^2(\beta)} = \\
\sqrt{cos^2(\beta)+a^2b^2\cdot sin^2(\beta)},
\label{eta_pol_rot_ellipse}
\end{split}
\end{equation}
the last equivalence is true because of the normalization condition given in Eq.(\ref{eta_pol_ellipse}), such condition also guarantees $\eta_{pol} \leq 1$, where the "equals" applies in the case $a=b=1$, which leads back to circular polarization for both beams, and to Eq.(\ref{eta_pol_circ}).
If $a=0$ or $b=0$, we are in the case of linear polarization for the two beams, and $\eta_{pol}=cos(\beta)$. 

\section{Measurement of the beams polarization states}
\begin{figure}[!ht]
    \centering
    \includegraphics[scale=0.3]{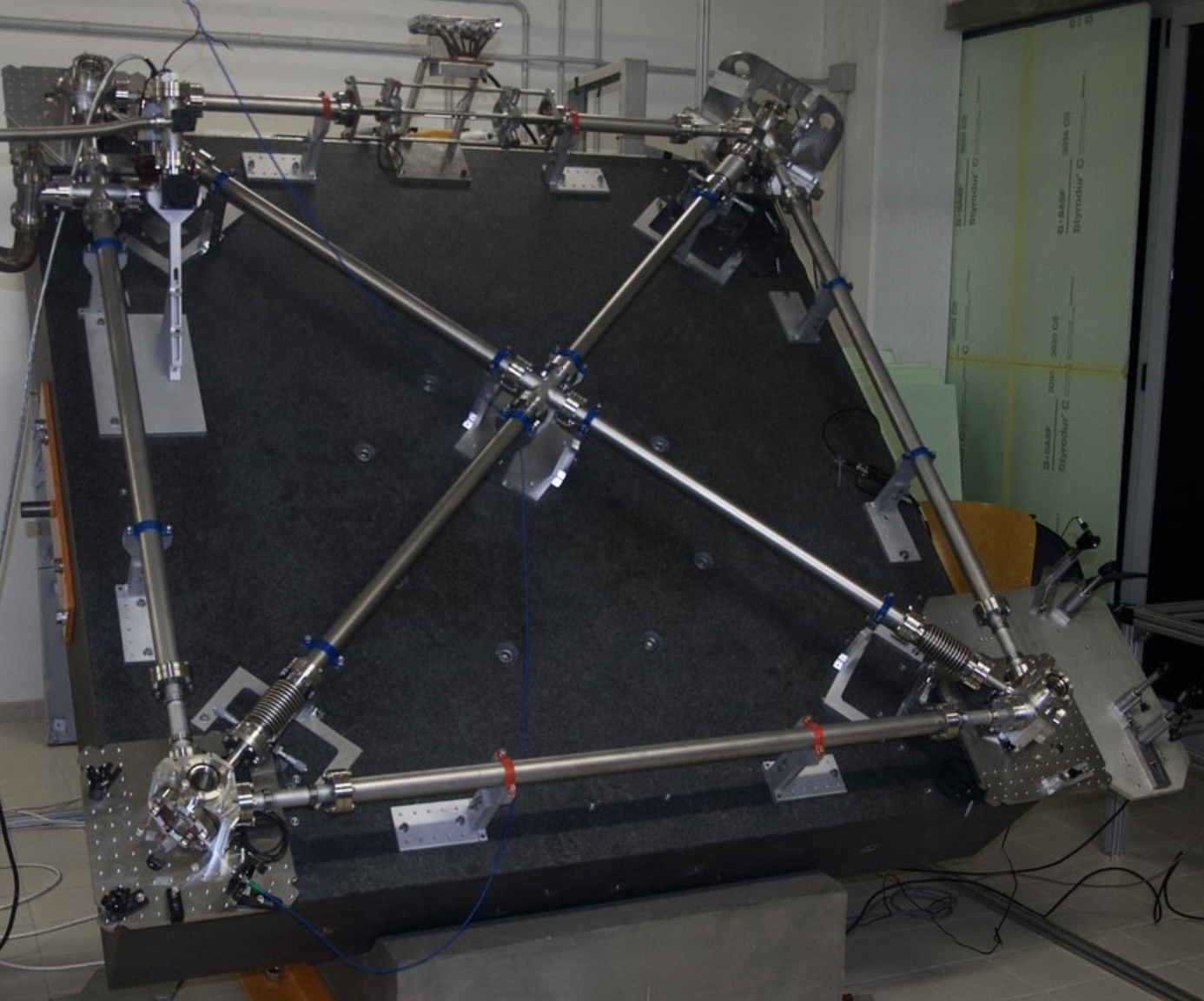}
    \caption{Experimental setup with "GP2" RLG at INFN-Pisa.}
    \label{fig:gp2}
\end{figure}

For GP2, observations of the output beams polarizations were performed, following a procedure similar to the one described in \cite{articolo_indiani}.
The RLG prototype is shown in Fig.\ref{fig:gp2}.

The whole setup is vacuum tight and filled with Helium and an isotopic 50/50 mixture of $^{20}$Ne and $^{22}$Ne. This choice allows two slightly different modes in the counter-propagating beams, in order to avoid mode competition \cite{virgilio2019}. The bias between the two modes is also given by the Earth rotation itself. The laser wavelength is $\sim 632.8$ nm, this results in a typical Sagnac frequency $f_s \simeq 184$ Hz, for this gyroscope. Mirrors are 1 inch diameter, 4m curvature radius, their transmission and birefringence properties are measured in the present work.

\begin{figure}[!ht]
    \centering
    \includegraphics[scale=0.3]{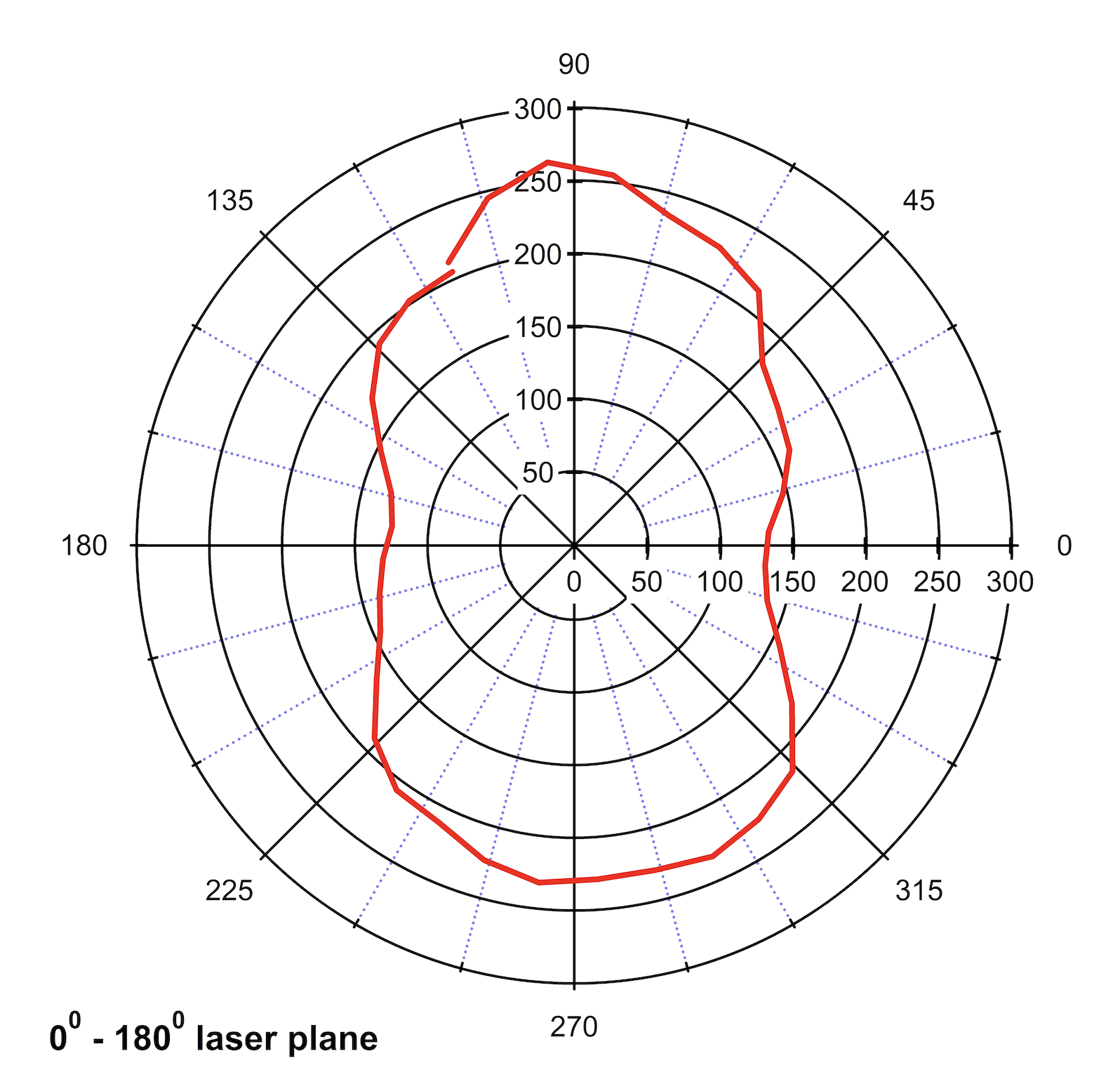}
    \caption{Polar graph obtained by the polarization analysis on one of the two laser beams, in GP2. Light intensity with respect to the polarizers' angle is  expressed in mV; $0^o$ and $180^o$ indicate the laser plane.}
    \label{fig:polar_graph}
\end{figure}

A full reconstruction of the polarization ellipse, from the intensity measured as a function of the polarizer's angle, was performed.
By means of a rotating polarizer and a photodiode, a polarization analysis has been performed on the CCW beam.
The resulting intensity polar graph (Fig.\ref{fig:polar_graph}) shows a peanut shape whose major to minor axes ratio is approximately 2:1, indicating an elliptical polarization with principal axes ratio 1.4:1, with a $\sim 10\%$ uncertainty.
The major axis of the polarization ellipse is ccw rotated of an angle $\phi^{'} = 44 \pm 17$ mrad with respect to the perpendicular to the laser plane. 

According to \cite{stedman90}, from the measurement of the beams' polarizations outside the cavity, it is possible to recover the polarization circulating in the ring laser cavity. First, with a dedicated experimental set-up we have measured the angular birefringence of a test mirror belonging to the same set of those mounted on GP2. A linearly polarized He-Ne laser beam is sent on the mirror at an angle of $45^{o}$ and the mirror induced polarization rotation on the reflected beam is analyzed by means of a polarizer. The measurement is performed both for vertical (S-type) and horizontal (P-type) laser polarization obtaining a birefringence $\chi = 11 \pm 1$ mrad. For the same test mirror, we measured the values of the intensity transmission coefficients $T_s$ and $T_p$ for S and P incident polarized light, giving for the fields amplitudes ratio $t_s/t_p = (T_s/T_p)^{1/2} = 0.062 \pm 0.002$, and resulting in an amplitude anisotropy $\delta = 90 \pm 10$ ppm.

Following \cite{stedman90}, when $\delta \ll \chi$ it is possible to calculate the out-of-plane misalignment angle $\alpha$ of the ring laser as:
\begin{equation}
\alpha \simeq \phi^{'} \cdot \chi \frac{t_s}{t_p} \cdot \sqrt{2} \simeq (43 \pm 15) \hspace{0.2 cm} \mathrm{\mu}\text{rad}
    \label{alpha}
\end{equation}

We can conclude that the polarization circulating inside the ring cavity is approximately linear, the electric field components ratio given by:
\begin{equation}
\frac{E_p}{E_s} = tan(\phi)  \simeq \phi = \frac{\alpha}{\chi} \frac{1}{\sqrt{2}} \simeq 0.0028 \pm 0.0003 \hspace{0.2 cm} \text{mrad}
    \label{e_ratio}
\end{equation}

\section{Interference of beams with linear and elliptical polarizations, measurements and comparison}

\subsection{Measurements of observables and comparisons}
\begin{figure}[!ht]
    \centering
    \includegraphics[scale=0.4]{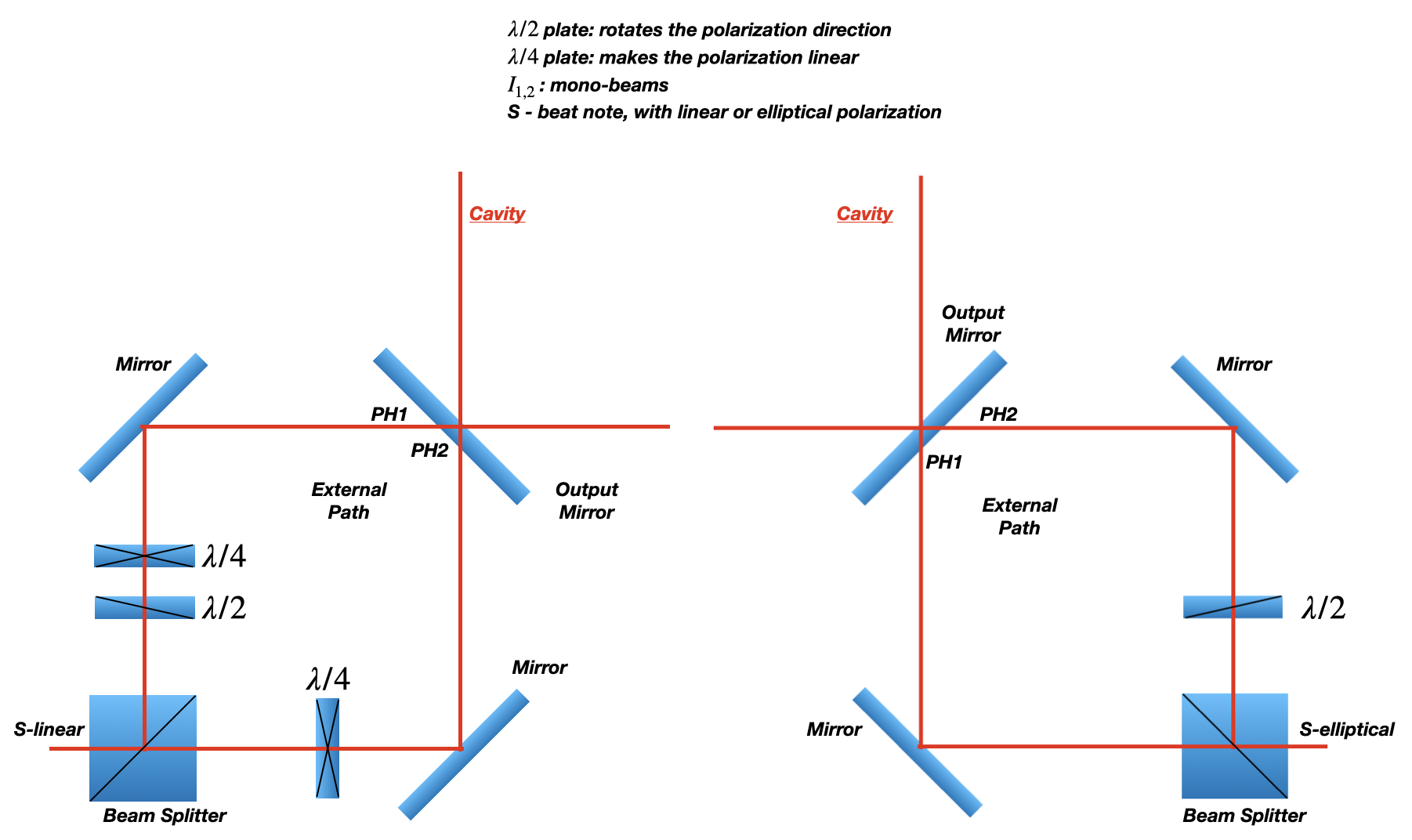}
    \caption{Schematic view of the 2 corners of the GP2 RLG, where two beat note signals are created by interfering beams with elliptical and linear polarization states.}
    \label{fig:schema_pol}
\end{figure}

The measurement is dedicated to understanding the impact of the interfering beams polarization on data quality; therefore, we compare the quality of the interference obtained with elliptically polarized beams outside the cavity (as measured and described in the previous paragraph) with quality obtained by making linear the polarizations of beams exiting the cavity.
Two distinct beat note signals were acquired, in the two lower corners of the square gyroscope: in one corner, the elliptic polarization states were rotated with respect to each other in order to maximize contrast, in the other corner the polarization states were made linear and then rotated in order to maximize the contrast; these operations were made by using $\lambda /2$ and $\lambda /4$ plates.
In Fig.\ref{fig:schema_pol} the creation of the two beat notes between linearly and elliptically polarized interfering beams is shown.

Data quality in the two acquisitions was evaluated on the basis of C and $\omega_m$.

\begin{figure}[!ht]
    \centering
    \includegraphics[width=4.0in]{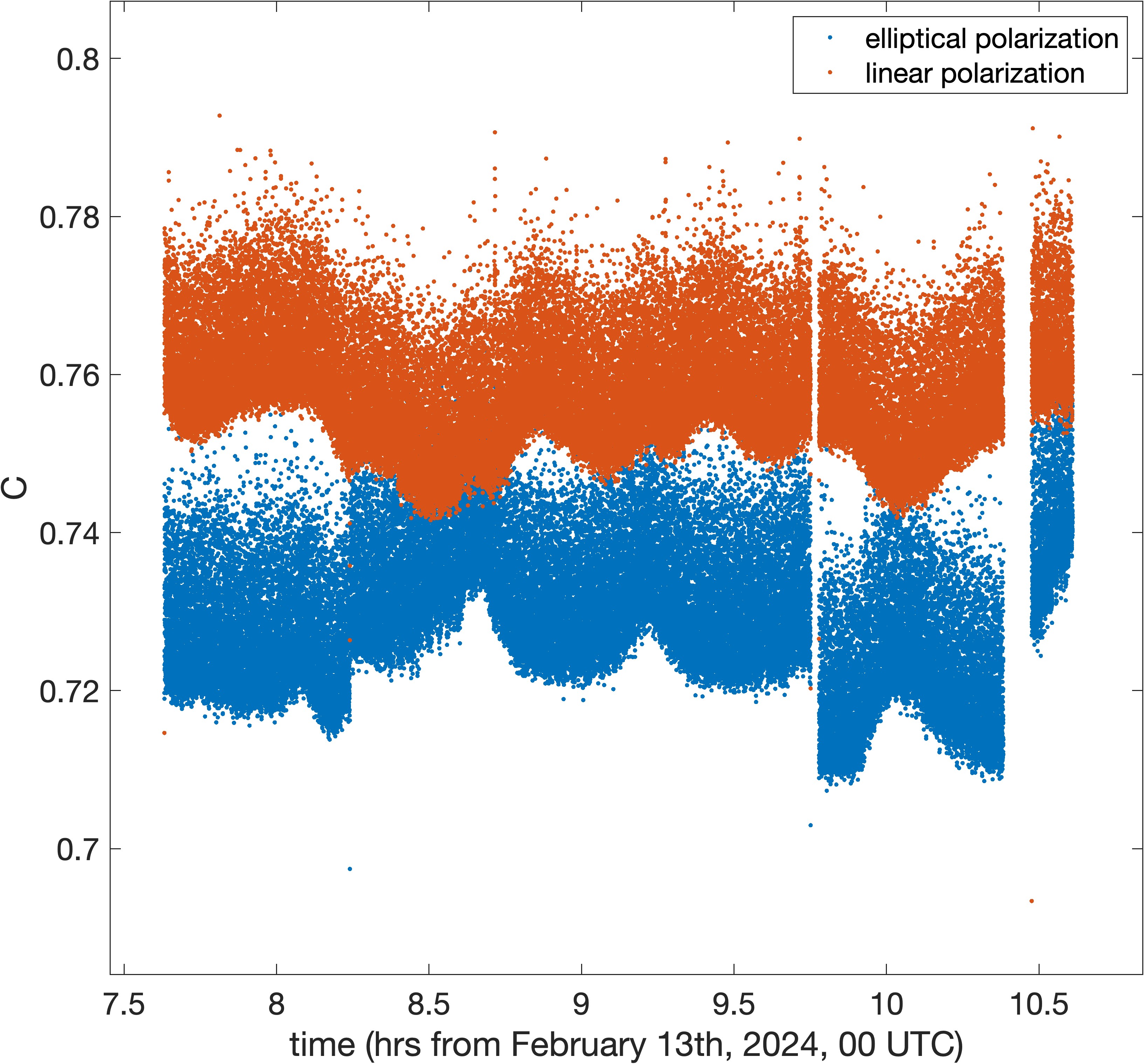}
    \caption{Contrast of the beat note acquired in two different corners of the RLG, with elliptically (blue) and linearly (orange) polarized interfering beams.}
    \label{fig:contrast_lin_ell}
\end{figure}

In Fig.\ref{fig:contrast_lin_ell}, we can see a comparison in contrast between beat note signals obtained with linearly and elliptically polarized interfering beams, on a portion of data chosen for its low noise level.

\begin{figure}[!h]
    \centering
    \includegraphics[width=4.0in]{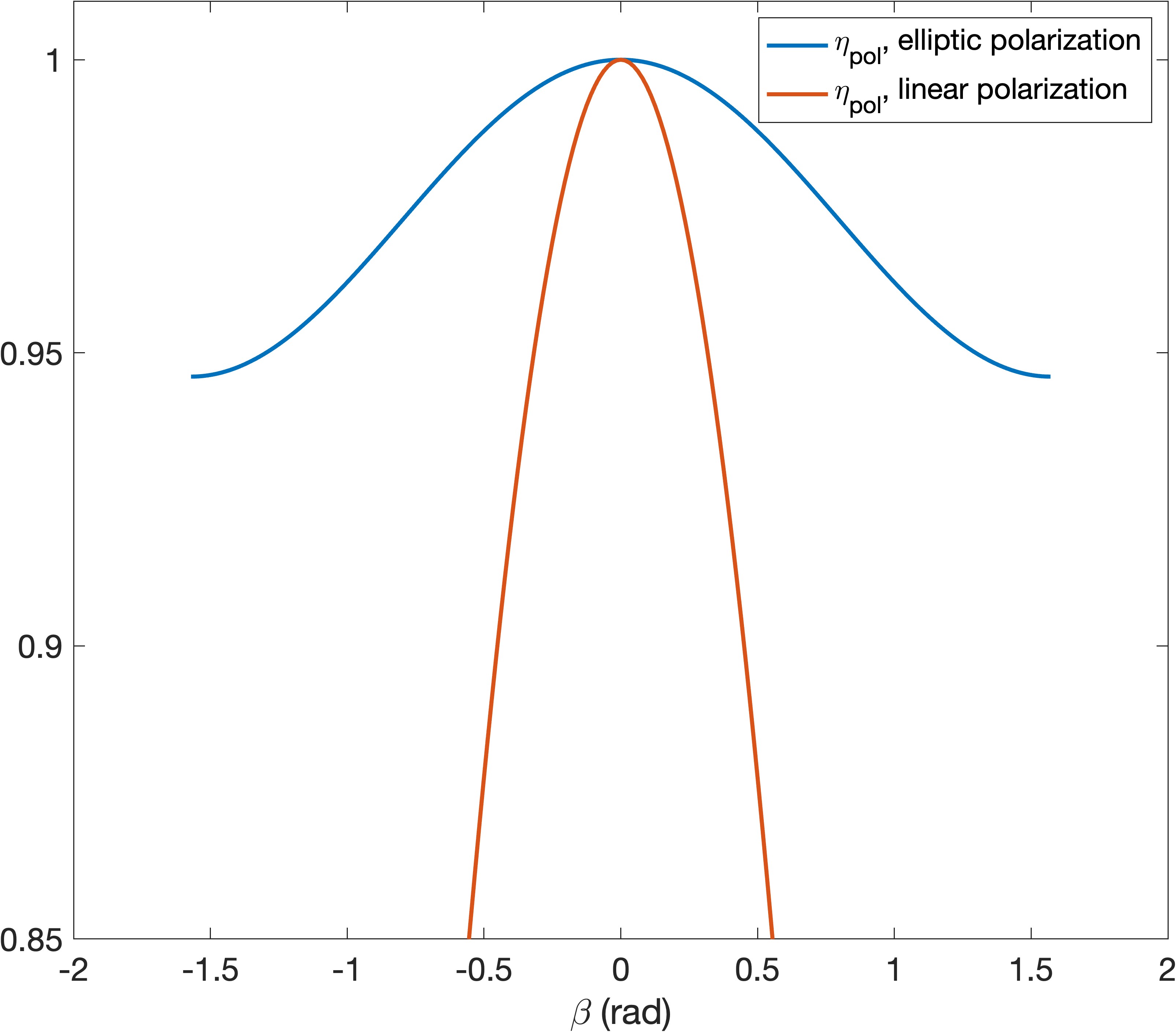}
    \caption{$\eta_{pol}$ vs $\beta$ is shown, for beams with linear and elliptic polarizations, with $\frac{a}{b}=1.4$.}
    \label{fig:eta_pol_lin_ell}
\end{figure}

The improvement, by obtaining an interference between linearly polarized beams, is evident. This can be explained by the model implemented for $\eta_{pol}$: 
in Fig.\ref{fig:eta_pol_lin_ell} $\eta_{pol}$ vs $\beta$ is shown, for beams with linear and elliptic polarizations, with $\frac{a}{b}=1.4$, in the second case, such ratio was measured between the axes of the polarization ellipse in GP2, as described in the previous paragraph.
As we can observe, in case of linear polarizations, the variation of $\eta_{pol}$ vs $\beta$ is much sharper, with respect to the case of "almost circular" polarization ellipses. 
This means that, by utilizing linearly polarized beams, we have a better sensitivity to the polarizations rotations and alignment, and as a result it is easier to obtain identical polarization states in interfering beams.

C and $\omega_m$ parameters show that we can obtain better performances from a RLG where the beams are made linearly polarized before interfering, as shown in table \ref{tab:parameters}. In particular, $\sigma(\omega_m)$ is an indicator of the signal-to-noise ratio.

\begin{table}
    \centering
    \begin{tabular}{|c|c|c|c|}
    \hline
     Polarizations & $\mu$(C) & $\sigma$(C) & $\sigma$($\omega_m/2\pi$)(Hz)\\
     \hline
        Linear & 0.761 & 0.0121 & 0.259\\
        Elliptic & 0.743 & 0.0134 & 0.265\\
    \hline
    \end{tabular}
    \caption{Parameters considered for comparing the performance of the RLG working with linearly or elliptically polarized interfering beams.}
    \label{tab:parameters}
\end{table}

\subsection{Measurement of the beams off axis}
\begin{figure}[!ht]
    \centering
    \includegraphics[width=4.0in]{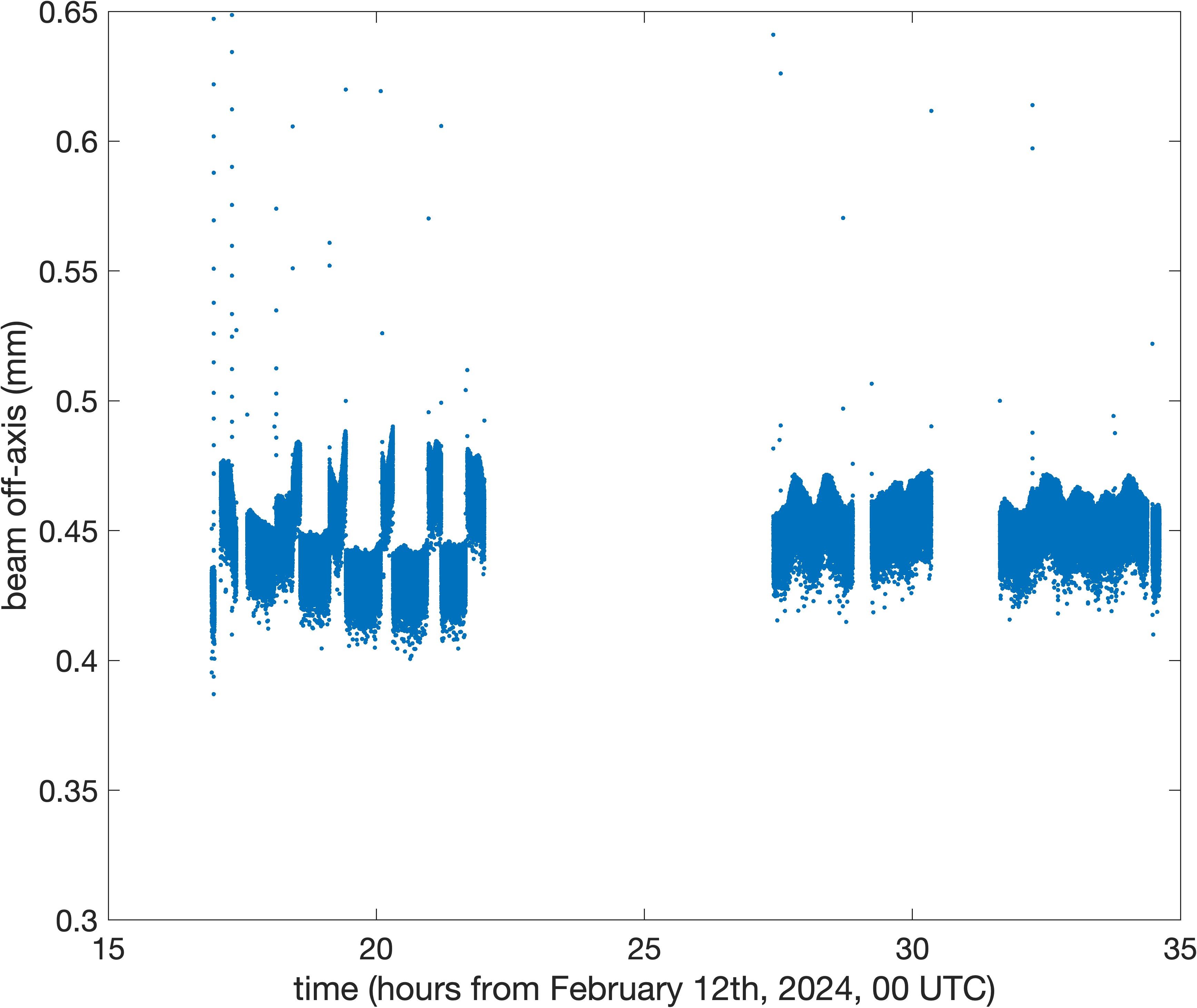}
    \caption{Beams off-axis in mm, estimated from linearly polarized beams interference}
    \label{fig:off_axis}
\end{figure}

Utilizing data taken with linearly polarized beams, the $\eta_{geom}$ factor of contrast was inferred, from Equations \ref{contrast_factors} and \ref{eta_vis}; we considered $\eta_{pol} \simeq 1$ in this case, and utilized the term including the beams intensities.
Utilizing the model described in Eq.(\ref{eta_geome_a}) and Fig.\ref{fig:eta_geome}, we first calculate the beams off-axis on selected "good data", in terms of the dimensionless parameter $k$, introduced in Paragraph 2.A, then we derive the beams off-axis.

The beam radius was measured by using a webcam Encore EN-WB-UHD01. The instrument registered the beam gaussian profile, and the result was $r=\sigma \cdot \sqrt{2}= 0.54 \pm 0.05$ mm, $r$ is the beam radius at which the beam intensities fall to $1/e^2$ of their axial values.
From the definition of $k$, we have, for the distance $d$, the beams off-axis:

\begin{equation}
d = 2 \cdot k \cdot r,
    \label{distance_d}
\end{equation}
such quantity is shown in Fig.\ref{fig:off_axis}.
Considering data taken on February $12^{th}$, 2024, and propagating the error on $k$ and $r$, we obtain for the measurement of the beam off-axis $d = 0.40 \pm 0.07$ mm, at 3$\sigma$ C.L.; it is reasonable to think the off-axis partially due to the cavity non-planarity estimated in Eq. (\ref{alpha}).

\newpage
\section{Conclusions}
A model of the Fringe Contrast in Ring Laser Gyroscopes beams interference was implemented; Contrast was factorized in terms of beams ratio, interfering beams alignment, and beams polarization.
The contributions of such factors were calculated, in particular the factors due to beams geometry and polarization were modeled and calculated for the case of GP2, a Ring Laser Gyroscope prototype at INFN-Pisa.
According to such model, GP2 was characterized, in its cavity planarity, beams alignment and polarization.
Beams polarization was measured at their interference point, and inside the cavity. The beams polarization at their interference point is elliptic with axes ratio $\simeq 1.4$, while inside the cavity it is approximately linear, the electric field components ratio given by:
$\frac{E_p}{E_s} \simeq$ 0.0028.
From such measurements, the cavity non-planarity angle was estimated to be $\alpha \simeq (43 \pm 15) \hspace{0.2 cm} \mathrm{\mu}\text{rad}$.

A measurement of Contrast with respect to the beams polarization was conducted, with a dedicated setup: on one corner of the Ring Laser, output beams polarization was made linear, while on the other corner it was left in its initial elliptical state.
Measurements of the Contrast were taken simultaneously with linearly and elliptically polarized beams, and compared. As a result, operating an RLG with linearly polarized interfering beams appears to be more convenient, in terms of Fringe Contrast, and signal-to-noise ratio. A measurement of the beams off-axis $d$ was also performed, exploiting the other factors contributing to Contrast; the result obtained gave $d = 0.40 \pm 0.07$ mm, at 3$\sigma$ C.L.

\section{Disclosures}
The authors declare no conflicts of interest.

\section{Funding} We acknowledge the "Istituto Nazionale di Fisica Nucleare" for funding the experiment.


\begin{thebibliography}{12}

\bibitem{lavoro_sismologia} A. Basti et al., "GINGER data analysis for seismology", ANNALS OF GEOPHYSICS, 67, 3, S320 2024, doi: 10.4401/ag-9033

\bibitem{belfi2017}J.Belfi et al., Deep underground rotation measurements: GINGERino ring laser gyroscope in Gran Sasso, Rev Sci Instrum 88, 034502 (2017), https://doi.org/10.1063/1.4977051

\bibitem{igel2021}H.Igel et al., ROMY: a multicomponent ring laser for geodesy and geophysics, Geophys. J. Int. (2021) 225, 684–698, doi:10.1093/gji/ggaa614

\bibitem{simonelli2016}A.Simonelli et al., First deep underground observation of rotational signals from an earthquake at teleseismic distance using a Large Area Ring Laser Gyroscope, Annals of Geophysics, https://doi.org/10.4401/ag-6970

\bibitem{simonelli2021}Simonelli, A.; Desiderio,
M.; Govoni, A.; De Luca, G.;
Di Virgilio, A. Monitoring Local
Earthquakes in Central Italy Using 4C
Single Station Data. Sensors 2021, 21,
4297. https://doi.org/10.3390/
s21134297

\bibitem{altucci2023status}C Altucci et al., Status of the GINGER project, AVS Quantum Sci. 5, 045001 (2023)

\bibitem{altucci2023memocs}C Altucci et al., GINGER, Mathematics and Mechanics of Complex Systems 11, 203–234 (2023)

\bibitem{capozziello2021}S. Capozziello et al., Constraining Theories of Gravity by GINGER experiment, Eur. Phys. J. Plus 136, 394 (2021)

\bibitem{giovinetti2024}F Giovinetti et al., GINGERINO: A high sensitivity Ring Laser Gyroscope for fundamental and quantum physics investigation, Front. Quantum Sci. Technol. 2024.1363409, (2024)

\bibitem{divirgilio2024}A. D. V. Di Virgilio et al., Noise Level of a Ring Laser Gyroscope in the Femto-Rad/s Range, Phys. Rev. Lett. 133, 013601 (2024)

\bibitem{fundamental} Di Somma, G.; Altucci, C.; Bajardi, F.; Basti, A.; Beverini, N.; Capozziello, S.; Carelli, G.; Castellano, S.; Ciampini, D.; De Luca, G.; et al. Possible Tests of Fundamental Physics with GINGER. Astronomy 2024, 3, 21–28. https://doi.org/10.3390/ astronomy3010003

\bibitem{wetzell} Di Virgilio, A.D.V., Terreni, G., Basti, A. et al. Overcoming 1 part in 10$^9$ of earth angular rotation rate measurement with the G Wettzell data. Eur. Phys. J. C 82, 824 (2022). https://doi.org/10.1140/epjc/s10052-022-10798-9

\bibitem{virgilio2019} A. Di Virgilio, N. Beverini, G. Carelli D. Ciampini, F. Fuso, E. Maccioni, Analysis of ring laser gyroscopes including laser dynamics, Eur. Phys. J. C  79: 573 (2019) https://doi.org/10.1140/epjc/s10052-019-7089-5

\bibitem{virgilio2020} A.D.V. Di Virgilio, N. Beverini, G. Carelli, D. Ciampini, F. Fuso, U. Giacomelli, E. Maccioni, A. Ortolan: Identification and correction of Sagnac frequency variations: an implementation for the GINGERINO data analysis, Eur. Phys. J. C 80:163 (2020) https://doi.org/10.1140/epjc/s10052-020-7659-6

\bibitem{articolo_recente_enrico} E.Maccioni et al., High sensitivity tool for geophysical applications: a geometrically locked ring laser gyroscope, Applied Optics Vol. 61, Issue 31, pp. 9256-9261 (2022) https://doi.org/10.1364/AO.469834

\bibitem{stedman90} H.R. Bilger, G.E. Stedman and P.V. Wells, "Geometrical dependence of polarisation in near-planar ring lasers", OPTICS COMMUNICATIONS (1990) Vol.80, n.2, Pages 133-137

\bibitem{articolo_polarizzazioni} L.Z. Cai, X.L. Yang, "Interference of circularly polarized light: contrast and application in
fabrication of three-dimensional periodic microstructures", Optics \& Laser Technology 34 (2002) 671 – 674 

\bibitem{articolo_indiani} Ramonika Sengupta, Brijesh Tripathi, and Asha Adhiya, "Explicit Reconstruction of Polarization Ellipse using Rotating Polarizer", arXiv:2211.15244



\end{thebibliography}
\end{document}